\documentstyle[12pt]{article}
\title{Lifschitz tail in a magnetic field: coexistence of classical
and quantum behavior in the borderline case.}
\author{L\'aszl\'o Erd\H os\thanks{Partially supported 
by the NSF grant DMS-9970323} \\School of Mathematics \\
 Georgia Institute of Technology \\
Atlanta, GA 30332\\
E-mail: {\verb -lerdos@math.gatech.edu-}}
\date{March 20, 2000}
\begin{document}
\baselineskip=.28in
\maketitle

\begin{abstract}
We establish
 the exact low-energy asymptotics of the integrated density of
states (Lifschitz tail) in a homogeneous magnetic field and 
 Poissonian impurities with a repulsive single-site potential
of Gaussian decay. It has been known that the Gaussian potential tail
discriminates between the so-called "classical" and "quantum"
regimes, and precise asymptotics are known in these cases.
For the borderline case, the coexistence of the classical and quantum
regimes was conjectured. Here we settle this last remaining
open case to complete the full picture
of the magnetic Lifschitz tails.

\end{abstract}


\medskip\noindent
{\it Running title:} Magnetic Lifschitz Tail: Borderline Case.


\newtheorem{theorem}{Theorem}[section]
\newtheorem{proposition}[theorem]{Proposition}
\newtheorem{corollary}[theorem]{Corollary}
\newtheorem{lemma}[theorem]{Lemma}
\newtheorem{definition}[theorem]{Definition}
\newtheorem{remark}[theorem]{Remark}

\newcommand{\sfrac}[2]{{\textstyle \frac{#1}{#2}}}
\newcommand{\rd}{{\rm d}}
\newcommand{\ov}{\overline}
\newcommand{\be}{\begin{equation}}
\newcommand{\ee}{\end{equation}}
\newcommand{\cE}{{\cal E}}
\newcommand{\bg}{{\bf g}}
\newcommand{\bal}{{\underline\alpha}}
\newcommand{\boldf}{{\bf f}}
\newcommand{\bE}{{\bf E}}
\newcommand{\bR}{{\bf R}}
\newcommand{\bP}{{\bf P}}
\newcommand{\bZ}{{\bf Z}}
\newcommand{\bL}{{\bf L}}
\newcommand{\bC}{{\bf C}}
\newcommand{\bM}{{\bf M}}
\newcommand{\bT}{{\bf T}}
\newcommand{\bH}{{\bf H}}
\newcommand{\Om}{\Omega}
\newcommand{\cA}{{\cal A}}
\newcommand{\cH}{{\cal H}}
\newcommand{\cI}{{\cal I}}
\newcommand{\cT}{{\cal T}}
\newcommand{\cG}{{\cal G}}
\newcommand{\cB}{{\cal B}}
\newcommand{\cC}{{\cal C}}
\newcommand{\cS}{{\cal S}}
\newcommand{\cR}{{\cal R}}
\newcommand{\cP}{{\cal P}}
\newcommand{\cQ}{{\cal Q}}
\newcommand{\cU}{{\cal U}}
\newcommand{\cK}{{\cal K}}
\newcommand{\cN}{{\cal N}}
\newcommand{\e}{\varepsilon}
\newcommand{\1}{{\bf 1}}
\newcommand{\tO}{\tilde\Omega}
\newcommand{\eps}{\varepsilon}
\newcommand{\om}{\omega}
\newcommand{\wt}{\widetilde}
\newcommand{\ua}{{\underline{\alpha}}}
\newcommand{\wh}{\widehat}




\section{Introduction}\label{intro}

The magnetic Lifschitz tail is the asymptotic behavior
of the integrated density of states (IDS), $N(E)$, at energy $E$
 near the bottom
of the spectrum of the two dimensional random Schr\"odinger operator
with a constant magnetic field $B$.
The random potential, $V_\om$,
represents repulsive impurities that are modelled
by a single-site potential profile $V^{(0)}\ge 0$
convolved with a homogeneous Poisson point process.

The low energy asymptotics of the IDS exhibits two qualitatively
different behaviors. For long range $V^{(0)}$, the asymptotics of
$N(E)$ is solely determined by the potential, i.e., by classical effects,
hence it is called {\it classical asymptotics} or {\it classical regime}.
In this regime, the low energy behavior of $N(E)$
 is sensitive to the details of the tail of  $V^{(0)}$
and it is insensitive to the strength of the magnetic field.
For short range potentials, the asymptotics
of $N(E)$ is determined by the quantum kinetic
energy ({\it quantum asymptotics} or {\it quantum regime}) and it
is universal; it depends only on the strength of the
magnetic field, but it is insensitive to  the potential
profile $V^{(0)}$. 

It has been established in \cite{BHKL} that potentials $V^{(0)}$ with
algebraically decaying tail of any finite order belong to the 
classical regime if $B\neq 0$.
The strength of the magnetic field does not appear in the leading
term asymptotics of $N(E)$. This result is in contrast
to the nonmagnetic case, where it has been shown (\cite{PF},
\cite{DV}, \cite{Sz}) that an algebraic decay,
$V^{(0)}(x)\sim |x|^{-(d+2)}$, discriminates between the classical
and quantum regimes in $d$ dimensions.

Nevertheless, quantum regime does appear in the magnetic case
as well, but the discriminating potential decay is much faster
than algebraic; in fact it is Gaussian. The existence of
the quantum regime for compactly supported potentials was
proven in \cite{E} and the Gaussian threshold was conjectured.
This threshold has been verified in \cite{HLW1} relying on \cite{E}
for the most involved technical part. More precisely, it has been
proven that {\it stretched-Gaussian} decay leads to the classical
asymptotics, while {\it super-Gaussian} decay leads to the quantum
asymptotics. Later, some refined results were obtained in
\cite{HLW2}.

The borderline case, when $V^{(0)}$ is asymptotically Gaussian,
has not been settled conclusively, only two-sided estimates were given
in \cite{HLW1}. The lower bound indicates a coexistence of the
classical and quantum effects, as it is determined by $2\ell_B^2 + \lambda^2$.
Here $\ell_B := B^{-1/2}$ is the magnetic lengthscale representing
the kinetic energy contribution,
 and $\lambda$ is the lengthscale of the Gaussian
potential. The upper bound is determined
by $\max\{ 2\ell_B^2, \lambda^2\}$, indicating  no coexistence
of the two regimes. The conjecture
of \cite{HLW1} was that the lower bound is the true asymptotics.

The purpose of this paper is to show this conjecture. We emphasize
 that in the Gaussian borderline case {\it both} classical
and quantum effects are important, hence none of them can be neglected
along the proof. This is the main novelty of the present paper,
which is a extension of our earlier work \cite{E}.

\subsection{Definitions}

We consider a nonnegative  potential function
\be
	V^{(0)} \in L^2_{loc}(\bR^2), \qquad V^{(0)} \ge 0 \; ,
\label{V1}
\ee
that is strictly positive on a non-empty open set, i.e.,
\be
	V^{(0)} (x) \ge v \cdot \1( |x-x_0|\le a)
\label{V2}
\ee
for some $v, a>0$ and $x_0\in \bR^2$.
Here $\1(\,\, \cdot\,\, )$ denotes the characteristic function.
 Let
\be
	V(x) = V_\om (x): = \sum_i V^{(0)} (x-x_i(\om))
\label{pot}
\ee
be a random potential, where $x_i(\omega)$ is the realization
of the Poisson point 
process on $\bR^2$ with a constant intensity
 $\nu$ (here $\omega$ refers to the randomness,
but we shall usually omit it from the notations). The expectation
with respect to this process
 is denoted by $\cE $.

We consider the following magnetic Schr\"odinger operator with
a random potential $V_\omega$
\begin{equation}
         H(B,V_{\omega})= H_{\omega} = \frac{1}{2} \Big[ 
        (-i\nabla - A)^2 - B\Big] + V_\omega \qquad \mbox{on}
	\quad L^2(\bR^2)\; ,
\end{equation}
where $A: {\bf R}^2 \rightarrow {\bf R}^2$ is a deterministic
vector potential (gauge) generating the constant $B>0$ magnetic field,
i.e., $\mbox{curl} \,\, A = B$.
The properties we are interested in
are independent of the actual gauge choice, so, conveniently, we
choose the standard gauge $A(x) := \frac{B}{2}{-x_2 \choose   x_1}$. 
Here $x=(x_1, x_2)\in\bR^2$. We subtracted the constant $B/2$ term
in the kinetic energy
both for physical reasons (spin coupling) and for mathematical
convenience. The spectrum of the free operator $H(B,V\equiv 0)$
is $\{ nB \, : \, n=0,1,2,\ldots\}$.

We also define $H_{Q,\om} = H_Q(B, V_\om)$ 
as the restriction of $H_\om$ 
onto a domain $Q \subset {\bf R}^2$ (with Dirichlet 
boundary conditions). In this paper, by domain we mean an open, bounded
subset of $\bR^2$ with regular (piecewise $C^1$) boundary,
 which is not necessarily connected.

We shall always assume that $V^{(0)}$ has  sufficient decay so that
$V_\omega \in L^2_{loc}$ with probability one, i.e.,
these operators  are almost surely selfadjoint.
Moreover, in all cases we consider it is easy
to show that
\be
	\inf \mbox{Spec} \; H(B, V_\om) = 0 \; \qquad \mbox{almost surely.}
\label{infspec}
\ee

We define the integrated density of states  (IDS) as
\begin{equation}
        N(E): = \lim_{Q \nearrow {\bf R}^2}
        \frac{1}{|Q |} \cE\, \mbox{Tr}\,  P_E(H_{Q,\om}),
\label{dens}\end{equation}
where $P_E$ is the spectral projection onto the half line $(-\infty , E]$,
and $Q\nearrow {\bf R}^2$ is an increasing sequence of nested
regular domains, say, squares or disks. The trace is over $L^2(Q)$.
For the existence of this limit and equivalent definitions we refer
to \cite{BHKL}, \cite{E} and references therein.

\bigskip

Following  \cite{HLW1},
 we assume that $V^{(0)}$
has one of the following behaviors at infinity:

{\bf Sub-Gaussian decay:}
$$
	\lim_{|x|\to\infty} {|x|^2\over \log V^{(0)}(x) } = -\infty \; ;
$$

\medskip

{\bf Gaussian decay:}
\be
	\lim_{|x|\to\infty}{\log V^{(0)}(x)\over |x|^2} = 
	- {1\over \lambda^2}
\label{gauss}
\ee
for some $0< \lambda< \infty$;

\medskip

{\bf Super-Gaussian decay:}
$$
	\lim_{|x|\to\infty}{\log V^{(0)}(x)\over |x|^2} = -\infty \; .
$$


The sub-Gaussian decay leads to the classical regime where
 the potential determines the Lifschitz tail. Hence precise results
require more definite tail behavior in this case. The following
definition is taken from \cite{HLW2}.

{\bf Regular $(F, \alpha)$-decay:}
$$
	\lim_{|x|\to\infty} {F(1/V^{(0)}(x))\over |x|}=1
$$
for some positive function $F$, which is regularly varying
of index $1/\alpha \in [0,\infty]$ and is strictly increasing
towards infinity.
 Recall that a positive measurable
function $F$ is said to be {\it regularly varying of index $\gamma$} if
$\lim_{t\to\infty} F(ct)/ F(t) = c^\gamma$
for all $c>0$. Such class of functions is denoted by $R_\gamma$.
Two  important cases are:

{\bf Algebraic decay:} $\lim_{|x|\to\infty} |x|^\alpha V^{(0)}(x) = \mu$
with some exponent $\alpha>2$ and constant $0<\mu <\infty$. This corresponds
to $F(t)\sim (\mu t)^{1/\alpha}$.

\medskip

{\bf Stretched Gaussian decay:}
$\lim_{|x|\to\infty} |x|^{-\alpha}\log V^{(0)}(x) =
- \lambda^{-\alpha}$
for some $0< \lambda< \infty$ and $0< \alpha < 2$.
This corresponds to $F(t) \sim \lambda (\log t)^{1/\alpha}$.
\medskip

\subsection{Results}\label{ressec}

The result of \cite{BHKL} for algebraically decaying potential $V^{(0)}$
is
$$
	\lim_{E\searrow 0} E^{2/(\alpha - 2)}\log N(E) = - C(\alpha, \mu,\nu)
$$
with an explicitly computed constant $ C(\alpha, \mu,\nu)$.

The  general regular $(F,\alpha)$-decaying
sub-Gaussian case was discussed in details in \cite{HLW2}.
The Lifschitz tail is given by the  de Bruijn conjugate $f^\#$
of the function $t\mapsto f(t) = [t^{-1/\alpha}F(t)]^{2\alpha/(2-\alpha)}$.
Recall that  the de Bruijn conjugate of a slowly
varying function $f \in R_0$ is $f^\#\in R_0$ such that
$f(t) f^\#(tf(t)) \to 1$ and $f^\#(t) f(tf^\#(t)) \to 1$
as $t\to\infty$.
With this definition 
$$
	\lim_{E\searrow0}{ E^{2/(\alpha - 2)} \log N(E)\over
	f^\#(E^{\alpha/(2-\alpha)})} = -C(\alpha, \nu) 
$$
with an explicit constant.
In particular,
for stretched-Gaussian potential $V^{(0)}$ this asymptotics is
explicitly  given as
(\cite{HLW1})
$$
	\lim_{E\searrow 0} {\log N(E)\over |\log E|^{2/\alpha}}
	= - \pi\nu\lambda^2 \; .
$$
For the super-Gaussian case it is proven in \cite{HLW1} that
\be
	\lim_{E\searrow 0} {\log N(E)\over |\log E|}
	= - 2\pi\nu\ell_B^2 = - {2\pi\nu\over B} \; ,
\label{supgauss}
\ee
with the additional assumption (\ref{V2}). 
In particular, the super-Gaussian decay includes compactly supported
potentials; the case for which (\ref{supgauss}) was proven in \cite{E}.
In \cite{HLW2} a slightly more general definition of
super-Gaussian decay was introduced:
\be
	\inf_{R>0} \mbox{ess}\sup_{|x|>R}{\log V^{(0)}(x)\over |x|^2}
	= -\infty \; ,
\label{gende}
\ee
and (\ref{supgauss}) was proven for such potentials (in addition to
the condition (\ref{V2})).

Finally, the following estimates were given in \cite{HLW1} for the 
Gaussian case (\ref{gauss})
\be
	-\pi\nu(\lambda^2 + 2\ell_B^2)
	\leq \liminf_{E\searrow 0} {\log N(E)\over |\log E|}
	\leq \limsup_{E\searrow 0} {\log N(E)\over |\log E|}
	\leq - \pi\nu\max \{ \lambda^2, 2\ell_B^2\} \; 
\label{twoside}
\ee
and the upper bound was weakened to $-2\pi\nu\ell_B^2$
in \cite{HLW2} if a more general definition of Gaussian
decay is used that is analogous to (\ref{gende}).
Our goal is to prove that the lower bound in (\ref{twoside})
is the correct one as conjectured in \cite{HLW1}.
\begin{theorem}\label{main}
Suppose that $V^{(0)}$ satisfies (\ref{V1}) and (\ref{gauss}).
Then 
\be
	\lim_{E\searrow 0} {\log N(E)\over |\log E|}
	= -\pi\nu(\lambda^2 + 2\ell_B^2), \qquad
	\ell_B:= B^{-1/2} \; .
\label{thmeq}
\ee
\end{theorem}

Since the lower bound (\ref{twoside}) has been proven in \cite{HLW1},
we focus only on the upper bound. As usual, we define the Laplace transform
of $N(E)$ as
$$
	L(t) : = \int_0^\infty e^{-E t}dN(E) = \cE e^{-tH_\om}(x,x) \; .
$$
Recall that 
the diagonal element of the averaged heat kernel is independent of $x$.
For more details, see \cite{E}.
Using a standard Tauberian argument (see for example Appendix
of \cite{HLW1} for details), the upper bound in
(\ref{thmeq}) follows from
\be
	\limsup_{t\to\infty} {\log L(t)\over \log t} \leq 
	 -\pi\nu(\lambda^2 + 2\ell_B^2) \; .
\label{lapl}
\ee

In the rest of the paper we prove (\ref{lapl}). Several steps
will be similar to \cite{E}, these will not be repeated in details.
 We give detailed proofs only for
the new parts of the argument.

\section{Localization}

We use a two-step localization as in \cite{E}. The
first  localization
is identical to the upper bound in Proposition 3.2 of \cite{E}
and the proof is the same.
\begin{proposition}\label{apri}
Let $M:=[-m, m]^2$ be a square box, then
$$
	L(t) \leq \liminf_{m\to\infty}
	 {1\over |M|} \cE \mbox{Tr} \; e^{-tH_{M, \om}} \; . \;\;\Box
$$
\end{proposition}

For the second, more refined localization we cannot
neglect the tail of the impurity potentials. We will define
effective {\it boundary potentials} that estimate
the potential tails inside a domain $\Omega$ that come
from impurities located outside of $\Omega$.

To prove (\ref{lapl}), it is enough to show that
\be
	\limsup_{t\to\infty} {\log L(t)\over \log t} \leq 
	 -\pi\nu(L^2 + 2\ell_B^2) \; 
\ee
for any $L<\lambda$. We fix two numbers, $0< L < \ov{L} <\lambda$, 
for the rest of the proof
and we omit the dependence on $L$ and $\ov{L}$ 
of various quantities in the notation.

Using (\ref{gauss}), there exists $R\ge 1$ such that
\be
	V^{(0)} (x) \ge e^{-|x|^2/\ov{L}^2} \qquad \mbox{for all} 
	\quad |x|\ge R \; .
\label{Vlow}
\ee
  We also
choose $R= R(\ov{L}, B)$ so large that $e^{-(2R)^2/\ov{L}^2} \le B$.
For any  domain $\Omega$ we  define
the following boundary potentials ($\partial\Omega$ stands for the
boundary of $\Omega$):
\be
	\ov{V}_\Omega (x) : 
	= \exp{\Big[ -(\mbox{dist}(x, \partial\Omega))^2/\ov{L}^2
	\Big]}
 	\cdot \1(x\in \Omega)\cdot
	\1 ( \mbox{dist}(x, \partial\Omega) \ge R)\; ,
\label{bpbar}
\ee
\be
	V_\Omega (x):=  
	\exp{\Big[ -(\mbox{dist}(x, \partial\Omega))^2/L^2
	\Big]}
 	\cdot \1(x\in \Omega) \; .
\label{bp}
\ee

\medskip

Similarly to Section 6 of \cite{E} we fix parameters
$0 < \beta <  B/2$, $1\leq s \leq m$ and let
$M:= [-m, m]^2$, $\wt M: = [-m-s, m+s]^2$,
$S:=[-s, s]^2$, $\wt S:= [-\sfrac{s}{2}, 
\sfrac{s}{2} ]^2$ and $Q_z:= Q+z$ for any square $Q\subset \bR^2$ and
 $z\in \bR^2$.
Finally, let $\lambda^{(B+2\beta)}_{S_z,\om}$ be
the lowest eigenvalue of
$$
	\ov{H}^{(B+2\beta)}_{S_z,\om} : = {1\over 2}
	\Bigg\{ \Big[-i\nabla - {B+2\beta\over 2} \pmatrix{-x_2\cr x_1}
	\Big]^2 - (B+2\beta )\Bigg\}
	+ V_\om + \overline{V}_{S_z}
$$
with Dirichlet boundary conditions on $S_z$. The magnetic field
of $\ov{H}^{(B+2\beta)}_{S_z,\om}$ is $B+2\beta$.
Notice that this operator differs from its counterpart in 
Section 6 of \cite{E}  by 
 the additional boundary potential $ \overline{V}_{S_z}$. We have

\begin{proposition}\label{secloc}
Assume that $\beta < 1/(2\ov{L}^2)$, 
 $\beta s^2 \ge 128$ and $s\ge 4R$. For any $z\in\bR^2$ there exists
a function $\eta_z$ supported on $S_z$ such that for any $f\in H_0^1(M)$
$$
	\langle f, H_{M,\om} f\rangle
	\ge {\beta\over 2\pi} \int_{\wt M}
	\rd z \langle f\eta_z, \ov{H}^{(B+2\beta)}_{S_z,\om}f\eta_z\rangle
	- 65s^{-2} e^{-\beta s^2/8} \| f\|_{L^2(M)} \; .
$$
\end{proposition}
Using this result, we obtain the following theorem from Proposition
\ref{apri} exactly as 
Theorem 6.3 was proven in
\cite{E}:
\begin{proposition}\label{secloccorr}
Let $\ell(t): = 10 (\log t / B)^{1/2}$, $s : = n_0\ell(t)$
and $S=[-s, s]^2$. For any fixed $ 0< \beta < 1/(2\ov{L}^2)$
 and $n_0\ge (B/\beta)^{1/2}$, $n_0\in \bZ$
\be
	\limsup_{t\to\infty} {\log L(t)\over \log t}
	\leq \limsup_{t\to\infty}(\log t)^{-1} \log \cE \exp
	\Big( - t \lambda^{(B+2\beta)}_{S,\om} \Big) \; . \qquad \Box
\label{Sest}
\ee
\end{proposition}

{\it Proof of Proposition \ref{secloc}.} Similarly to the proof
of Proposition 6.1 in \cite{E}, we define
$$
	\varphi_z(x): = e^{-\beta(x-z)^2/2}e^{i\beta (x_2z_1-x_1z_2)} \; 
$$
and $T_\beta: = -i\partial_1 + \partial_2 + (B/2 +\beta)x_2
	- i(B/2 + \beta)x_1$.
We use the following identity to localize the kinetic energy
for $f\in H^1_0(M)$
$$
	\langle f, H_{M,\om} f\rangle
	= {\beta\over \pi} \int \rd z \int \Big\{
	\sfrac{1}{2} |T_\beta (\varphi_z f)|^2 + V_\om |\varphi_z f|^2
	\Big\}\; ,
$$
where we let $\int$ denote $ \int_{\bR^2}$. This magnetic localization
principle was first used in \cite{ES}.

Fix a smooth function $\theta(x)$ such that $\theta \equiv 1$
on $\wt S$, $\theta \equiv 0$ on $\bR^2\setminus [-\sfrac{3s}{4}, 
\sfrac{3s}{4}]^2$, $0\leq \theta \leq 1$ and $\|\nabla\theta \|_\infty
\leq 8s^{-1}$. Let $\theta_z(x):= \theta(x-z)$ and $\eta_z:= \theta_z
\varphi_z$.
The function $\theta_z$ can be commuted
with $T_\beta$ at the expense of an error of size $\|\nabla \theta_z\|^2$
on the support of $\nabla\theta_z$. The result is
the analogue of (6.13) in \cite{E}
\be
        \langle f, H_{M ,\omega} f\rangle
        \geq {\beta\over 2\pi}\Bigg[\int_{\wt M_z} 
        \Bigl\{ \int \frac{1}{2} |T_\beta ( \eta_z f) |^2 
        + \int V_\omega|\eta_zf|^2 \Bigr\} \rd z
        - 128\pi \beta^{-1}s^{-2}  e^{-\frac{\beta}{8}s^2}\Bigg]\; .
\label{lq3}\ee 

Finally, we  estimate the boundary
 term $\int \rd z  \int \overline{V}_{S_z}  |\eta_z f|^2$
in  $\ov{H}^{(B+2\beta)}_{S_z,\om}$ using 
$\overline{V}_{S_z} |\eta_z f|^2\leq e^{-(s/4\ov{L})^2} |\varphi_z|^2
|f|^2$;
$$
	{\beta\over 2\pi}\int_{\wt M_z} \rd z  \int \ov{V}_{S_z}
	  |\eta_z f|^2
	\leq  e^{ -s^2/(4\ov{L})^2}	\int |f|^2 \; ,
$$
which can be included into the error term in (\ref{lq3}). $\;\;\Box$.

\section{Enlargement of obstacles}\label{enlarge}

We follow the basic strategy of Sznitman \cite{Sz}
and its magnetic version from \cite{E} to estimate the
lowest eigenvalue $\lambda^{(B+2\beta)}_{S,\om}$ of 
$\ov{H}^{(B+2\beta)}_{S,\om}$ by the lowest eigenvalue of
a Hamiltonian with enlarged, hard-core obstacles. 
We need this argument for $\ov{H}^{(B+2\beta)}_{S,\om}$, but
the actual  field does not play much role
in this section, so for brevity we consider
$ \ov{H}^{(B)}_{S,\om}$ and let $\lambda^{(B)}_{S,\om} $ 
be its smallest eigenvalue.

 The main novelty is that
we cannot simply use Dirichlet boundary conditions for the
"enlarged obstacle" Hamiltonian, since the potential tail
penetrating into the clearing regimes does influence
the lowest eigenvalue. We add an appropriate Gaussian boundary potential
to the hard-core Dirichlet wall, and we also keep the Gaussian
tail of the original potentials.
 The obstacle  configuration
$\om$ is fixed throughout this section.

The "enlarged obstacle" Hamiltonian requires several
definitions that were listed in Section 7.1 of \cite{E}.
Here we recall only that four parameters, $\ell$, $b$, $\eps>0$
and $r>0$ have to be fixed. 
With these parameters, one defines (Section 7.1 of \cite{E})
the set of "good" points
(their indices denoted by $\cG$), clearing boxes and the set $A^1$, which
is the $\ell$-neighborhood of clearing boxes. 
Recall that a point $x_i$ is "good" if it is not isolated
from other points in a certain hierarchical sense. Clearing boxes
are squares of size $\ell$ that contain a large regular set ("clearing")
free of good points.
 Finally we define, for $s>b$,
\be
	\Omega : = S \setminus \bigcup_{i\in \cG}
	\Big[ \overline{B}(x_i, 2R) \setminus B(x_i, R)\Big]\; ,
	\qquad
	\Omega_+^b:= \Big( [-s+b, s-b]^2\cap A^1\Big)
	\setminus \bigcup_{i\in \cG}\overline{B}(x_i, b) \; ,
\label{Omdef}
\ee
where $B(x, \rho)$ denotes the open ball of radius $\rho$ about $x$.
 We choose $\delta = 
\sfrac{R}{100}$. Notice that $\Omega$ is defined by removing
annuli around the good points, unlike in \cite{E}, where balls
were removed.  $\Omega_+^b$ is the "clearing set",
where the  "enlarged obstacle" Hamiltonian will be defined.

We let 
$$
	U(x): = e^{-|x|^2/\ov{L}^2} \cdot \1 (|x|\ge R) , \qquad
	\wt V_\om(x): =
	\sum_{i \in \cG} U(x-x_i(\om))\; ,
$$
then we clearly have $V_\om \ge \wt V_\om$.
 This definition of $\wt V_\om$ is different from (7.3) of \cite{E}.
The role of $v$ in \cite{E} will be played by the constant 
$e^{-(2R)^2/\ov{L}^2}$;
this is a lower bound on the potential $\wt V_\om$ in the annuli
$\{ R \leq |x-x_i|\leq 2R \}$ around the good points. The role of $a$
in \cite{E} is played by $2R$. The specific upper
bound $a\leq 1$ imposed in \cite{E} will not be important.

\medskip

We will estimate the lowest eigenvalue, $\wt\lambda$, of the Hamiltonian with
potential $\wt V_\om$ by  $\wt \lambda_b$, the lowest  
eigenvalue with  hard core
potential on $\Omega_+^b$. We add boundary potentials to both Hamiltonians.
 Since we work on multiply connected domains,
we must take the gauge freedom
into account as in Section 7.2 of \cite{E}. Hence both eigenvalues
are defined as the infimum over all gauges on the 
complementary domain of the obstacles.
We recall that for any
$\underline{\alpha}= \{ \alpha_i \}_{i\in \cG} \in [0, 2\pi)^\cG$
we defined $B_{\underline{\alpha}}(x): = B + \sum_{i\in \cG} \alpha_i
B^*(x-x_i)$ and its radial gauge $A_{\underline{\alpha}}$,
$\mbox{curl}\,\,  A_{\underline{\alpha}} = B_{\underline{\alpha}}$,
 where $B^*:= (4/\pi)\cdot \1_{B(0,1/2)}(x)$ (the definition
of $B^*$ in \cite{E} missed a ${1\over 2\pi}$ factor). The magnetic field 
$B_{\underline{\alpha}}$ includes flux tubes of strength $\alpha_i$
around the good points. We define
\be
	\wt \lambda =\wt \lambda(B) : = \inf_{\underline{\alpha}} 
	\lambda_{\underline{\alpha}} \; , \qquad
	\lambda_{\underline{\alpha}}: = 
	\inf \mbox{Spec}  \Bigg( {1\over 2} 
	\Big[ (-i\nabla -  A_{\underline{\alpha}})^2 -  B_{\underline{\alpha}}
	\Big] + \wt V_\om + \overline{V}_S\Bigg)_S \; ,
\label{tildelambda}
\ee
where the subscript refers to Dirichlet boundary conditions on $S$.
Clearly $\lambda_{S,\om}^{(B)}\ge \wt \lambda$. Similarly,
\be
	\wt \lambda_b=\wt \lambda_b(B) : 
	= \inf_{\underline{\alpha}} \lambda_{b,\underline{\alpha}}
	\; , \qquad
	\lambda_{b,\underline{\alpha}}: =\inf \mbox{Spec}\Bigg( {1\over 2} 
	\Big[ (-i\nabla -  A_{\underline{\alpha}})^2 -  B_{\underline{\alpha}}
	\Big]  + V_{\Omega^b_+} \Bigg)_{\Omega^b_+} \; ,
\label{tildelambdab}
\ee
again with Dirichlet boundary conditions on $\Omega^b_+$.
Notice that the decay of the boundary potential $ V_{\Omega_b^+}$
is slightly stronger than that of $\widetilde V_\om + \ov{V}_S$
since $L < \ov{L}$.

Let $g_U$ denote the Green's function of
any domain $U$, i.e., the solution to $\Delta g_U =-1$
on $U$ and $g_U=0$ on $\partial U$. We let
$G_U: = \max_{x\in \ov{U}} g_U(x)$. 

The importance of these functions is that
the lowest magnetic Dirichlet eigenvalue of  a large domain $U$ is
essentially $e^{-2BG_U}$ (a factor 2 was missing on page 349 of 
\cite{E}), and the eigenfunction is
roughly $e^{Bg_U}$ with some cutoff near the boundary.
Moreover, for   "round" domains, $g_U$
is roughly quadratic in the distance from the boundary.
Hence, roughly,
$$
	\wt\lambda_b \sim \exp{ (-2BG_{\Omega^b_+})} + \int_{\Omega^b_+}
	 \exp{ \Big( - \Big[ {\mbox{dist}(x, 
	\partial\Omega^b_+)\over L}\Big]^2 \Big)}
	\Bigg| {\exp{ Bg_{\Omega^b_+}(x)}\over
	\| \exp{ Bg_{\Omega^b_+}} \|} \Bigg|^2 \rd x \; .
$$
Here the first term represents 
the kinetic energy due to localization in the clearing.
The second term is the interaction of the "quantum"
wavefunction with the "classical" effect; the effective
contribution of the potential tails. It turns out that the
second term dominates. The main contribution
comes from the interplay between the Gaussian character of the
magnetic eigenfunction
$\approx \exp{ Bg_{\Omega^b_+}}$ and the Gaussian potential.

\bigskip

The basic comparison result is the analogue of Corollary 7.3 
 in \cite{E} (there  are two misprints in (7.16) in \cite{E};
 $r\to\infty$  should be $r\to0$ and a minus sign is missing in
front of $\log K$).

\begin{proposition}\label{compprop}
For any fixed positive integer $n_0$ we let $s:=n_0\ell$ and let
$\varrho>0$ be a positive number.
For small enough $r$, $\varepsilon$,
there exist  $K= K(b, B, r, \ell, s, L, \ov{L}, \varepsilon, \varrho)$ and 
$w(r)$ with $\lim_{r\to 0} w(r) =1$ such that 
 $\wt\lambda_b^{w(r)}
\leq \wt \lambda/K$ if $\wt\lambda\leq \min\{ 4K,e^{-\varrho BG_\Omega}\}$,
 and $K$ satisfies
\be
	\limsup_{r\to0}\limsup_{b\to\infty\atop \eps\to0}
	\limsup_{\ell\to\infty} {-\log K\over \ell^2} = 0 \; .
\label{K1}
\ee
\end{proposition}

The basic intuition behind this comparison is that if the lowest
eigenvalue of $H_{S, \om}$ is very small, then there must be
a big clearing  in the obstacle configuration, and the lowest
eigenfunction is essentially supported in this clearing. 
Hence
this eigenvalue can be estimated by the Dirichlet eigenvalue
within the clearing even with enlarged obstacles.
 The inclusion of the boundary potential 
does not change this mechanism, but it changes both eigenvalues.
The threshold for such eigenvalues is controlled by two
different functions. The control given by $K$ is analogous
to \cite{E}. The control $\wt\lambda\leq e^{-\varrho BG_\Omega}$
is new.

\bigskip

The proof of Proposition \ref{compprop} is similar to that of
Theorem 7.2 in \cite{E}, but we have to include the boundary potential.
We first show that the increase of the eigenvalue due to the
enlargement is given by the size of the eigenfunction near the
boundary (Lemma \ref{near}). Then, by applying  a probabilistic
argument, we show that $g_\Omega(x) \ll G_\Omega$ if $\Omega$ is large
and $x$ is close to the boundary (Lemma \ref{gG}). 
In other words, the eigenvalue increases by at most  a factor
$e^{o(BG_\Omega)}$.  For technical reasons we
give these estimates for a slightly enlarged domain 
$$
	\Theta :  = \Omega + B(0, 2\delta) \; . 
$$

In \cite{E} (Lemma 7.7), we finally
 estimated $G_\Omega$ by the logarithm of the
 magnetic Dirichlet eigenvalue of
$\Omega$ to show that $e^{o(BG_\Omega)}\leq \wt\lambda^{-o(1)}$
and therefore $\wt\lambda_b \le \wt\lambda^{1-o(1)}$.
The analogue of this estimate with a boundary potential
is more complicated because it requires a  control on $g_\Omega$
not only near the boundary.
But fortunately we do not need this estimate with a precise constant
since it is used only in the error factor $e^{o(BG_\Omega)}$.
So we choose an alternative
method that estimates $G_\Omega$ by the logarithm of the
 magnetic Dirichlet eigenvalue without boundary potential,
exactly as in \cite{E}. The new control 
$\wt\lambda\leq e^{-\varrho BG_\Omega}$ stems from this modification.

 We will state these lemmas precisely,
but we give details of the proof
 only for the modifications compared with \cite{E}.

\begin{lemma}\label{near}
There exist positive 
constants, $c_1, c_2$, depending only on $B, L, \ov{L}, R, b$, such that
\be
	 \wt \lambda_b \leq c_1 \wt\lambda
	 s^2 e^{2B\eta} \; ,
\label{neareq}
\ee
whenever $\wt \lambda \leq c_2 s^{-2} e^{-2B\eta}$, $s> 2b\ge 40R$, where
$$
	\eta  = \max\Big\{ g_{\Theta}(z) \; : \;
	z\in \overline{\Theta}\setminus \Omega_+^{2b}
	\Big\}\; , \qquad
	\Theta:= \Omega + B(0, 2\delta) \; ,  \qquad \delta:={R\over 100} \; .
$$
\end{lemma}

{\it Proof.} We fix $\underline{\alpha} \in [0, 2\pi)^\cG$ and let
$\varphi_{\underline{\alpha}}$ be the  normalized
eigenfunction belonging 
to $\lambda_{\underline{\alpha}}$. We can assume
that $\lambda_{\underline{\alpha}} \leq c_2 s^{-2} e^{-2B\eta}$.

Let $T_{\underline{\alpha}}: = -i\partial_1 +\partial_2 - 
(A_{\underline{\alpha}})_1 - i (A_{\underline{\alpha}})_2$, then
by variational principle and integration by parts
$$
	\lambda_{b, \ua} = \inf_{\psi\in H_0^1(\Omega_+^b)}
	{ \int_{\Omega_+^b}\sfrac{1}{2} | T_\ua \psi|^2
	+ V_{\Omega_+^b} | \psi|^2 \over
	\int_{\Omega_+^b} |\psi|^2} \; .
$$
Let $\theta$ be a cutoff function such that $\theta\equiv 1$
on $\Omega_+^{2b}$, $\theta\equiv 0$ on $\bR^2\setminus\Omega_+^{b}$,
$0\leq \theta\leq 1$ and $|\nabla\theta|\leq 4b^{-1}$.
Then
$$
	\lambda_\ua = {1\over 2}\int_S |T_\ua \varphi_\ua|^2
	+\int_S (\wt V_\om + \ov{V}_S)|\varphi_\ua|^2
$$
$$
	\ge {1\over 4} \int_S 
	|T_\ua (\theta\varphi_\ua)|^2 - \| \nabla \theta\|^2_\infty
	\int_{S\cap supp(\nabla \theta)} 
	|\varphi_\ua|^2 + c_3^{-1}\int_S V_{\Omega_+^b}|\varphi_\ua|^2 \; ,
$$
using the pointwise inequality $  V_{\Omega_+^b}(x)
\leq c_3\big[ \wt V_\om (x)+ \overline{V}_S(x)\big]$ with
some $c_3 = c_3(B, L, \ov{L}, b)\ge 1$. We use $\psi:=\theta\varphi_\ua$
as a trial function to obtain
$$
	\lambda_{b, \ua} \leq  \, {2c_3\lambda_\ua + 16b^{-2} 
	\int_{S\setminus\Omega_+^{2b}} |\varphi_\ua|^2\over
	1 - \int_{S\setminus\Omega_+^{2b}} |\varphi_\ua|^2}
$$
similarly to (7.19) in \cite{E}.

To complete the  proof of (\ref{neareq}), we need the upper estimate  
$$
	\int_{S\setminus\Omega_+^{2b}} |\varphi_\ua|^2
	\leq c_4\lambda_\ua s^2 e^{2B\eta}
$$
(with some $c_4= c_4(B,\ov{L},R,b)$)
 whose derivation is identical 
to the rest of the proof of Lemma 7.4 \cite{E}.
The only difference is that the balls $\overline{B}(x_i, a)$ are
replaced with the annuli $\overline{B}(x_i, 2R) \setminus B(x_i, R)$,
according to the new definition of $\Omega$ in (\ref{Omdef}),
and the constant $v$ is replaced with $e^{-(2R)^2/\ov{L}^2}$.
In particular
$$
	S\setminus \Omega_+^{2b} \subset
	\Big( \Omega\setminus \Omega_+^{2b}\Big)
	\cup  \bigcup_{i\in \cG}
	\Big[ \overline{B}(x_i, 2R) \setminus B(x_i, R)\Big] \; ,
$$
and we estimate 
$$
	\int_{ \bigcup_{i\in \cG}
	\big[ \overline{B}(x_i, 2R) \setminus B(x_i, R)\big]}
	|\varphi_\ua|^2 \leq \lambda_{\ua} e^{(2R)^2/\ov{L}^2}
$$
instead of the second inequality in (7.20) in \cite{E}.
The details are omitted.
$\;\;\;\Box$

The next lemma is an analogue of Lemma 7.5 in \cite{E}.
It states that near the boundary of the enlarged obstacles
the Green's function $g_\Theta$ is 
much smaller than its maximum $G_\Theta$.

\begin{lemma}\label{gG}
Let $20R\leq b$, $40b\leq \ell \leq s$, $r< 1/4$ and
$G_\Theta \ge c_5\ell$ with some $c_5 = c_5(R)$.

(i) For small enough $\varepsilon$, there exist 
 $\ell_0(\e, b)>0$
and $0< k=k(\e, b)\leq 1/4$ 
such that
\be
	\sup_{x\in \Sigma\cap\Theta }
	g_\Theta(x) \leq \Big({b \over \ell}\Big)^{k} G_\Theta
	 \qquad \mbox{for} \quad
	\ell\ge\ell_0(\e, b)
\label{onsigma}
\ee
with $\Sigma: = \Big( \ov{S}^{2\delta}_- \setminus S_+^{2b}
\Big) \cup \bigcup_{i\in \cG} \overline{B}(x_i, 2b)$, where
$S_\pm^c: = (-s\pm c, -s\mp c)^2\subset\bR^2$.

(ii) There exists a positive number $c_0$ such that
\be
	g_\Theta(x) \leq \Big[ (1-c_0)^{1/r} G_\Theta + c_0^{-1}r^2\ell^2
	\Big] + \sup_{y\in \Sigma\cap \Theta} g_\Theta(y)
\label{inforest}
\ee
for all $x\in \Theta$, $x\not\in A^1\cap S$.
\end{lemma}

{\it Proof.} The proof is almost identical to that of Lemma 7.5
\cite{E}. The only difference is that $\Omega$ is defined by
removing annuli,
 hence the case $x\in B(x_i, R+2\delta)$, $i\in \cG$,
needs a separate estimate. For such $x$, the exit time from 
$\Theta$ is at most the hitting time of the circle
 $\partial B(x_i, R+2\delta)$, hence its expected value 
depends only on $R$. This estimate is taken into account in 
Lemma \ref{gG} by the extra requirement $G_\Theta \ge c_5\ell$. $\;\;\;\Box$

We use the following lemma to establish that the maximal 
expected hitting time is essentially the same for $\Omega$
and $\Theta = \Omega + B(0, 2\delta)$. The proof is identical
to that of Lemma 7.6 \cite{E}; the geometric condition
used in \cite{E} is satisfied for the new definition of $\Omega$ (\ref{Omdef})
as well.
\begin{lemma}
There exists two constant $c_6, c_7$ depending on $R$ such that
\be
	G_\Omega \leq G_\Theta \leq c_6 G_\Omega + c_7 \; . \qquad \qquad 
	\Box
\label{Omcomp}
\ee
\end{lemma}

Now we are ready to prove Proposition \ref{compprop}. We note
that $\overline{\Theta} \setminus \Omega_+^{2b} \subset \Sigma \cup (A^1)^c$.
Hence for large enough $\ell$,
 the combination of (\ref{onsigma}), (\ref{inforest}) and
(\ref{Omcomp}) gives
\be
	\eta \leq c_8\Big[ (1-c_0)^{1/r} G_\Omega + c_0^{-1}r^2\ell^2
	\Big]
\label{etabecs}
\ee
with some $c_8=c_8(R)$, similarly to (7.43) in \cite{E}. 
This means that $\eta \ll G_\Omega$ for small $r$  if
$G_\Omega\gg r^2\ell^2$.
Since $\wt\lambda \leq e^{-\varrho BG_\Om}$, we see that $e^{2B\eta}$ is
bounded by a small inverse power of $\wt\lambda$, so
from (\ref{neareq}) we get that $\wt\lambda_b \leq \wt\lambda^{1-o(1)}$
as $r\to0$. The case $G_\Omega = O(r^2\ell^2)$ can be included
in the error factor $K$.
The details are very similar to \cite{E} and are left to the reader.
$\,\,\,\Box$.

\section{Proof of the upper bound in Theorem \ref{main}}

We recall the definition of $\wt\lambda(B)$ and $\wt\lambda_b(B)$ from
(\ref{tildelambda}) and (\ref{tildelambdab}). Using
 the notations and results of  Section \ref{enlarge} for $B$ replaced
with $B+2\beta$, we have
 $\wt\lambda(B+2\beta)\leq \lambda_{S,\om}^{(B+2\beta)}$ with
$0\leq \beta \leq B/2$,  $s=n_0\ell(t)$ 
and $\ell=\ell(t) = 10\sqrt{{\log t\over B}}$.
We will need $\Omega$ defined in (\ref{Omdef}) and we note that
 $\Omega$ depends on $B$, $R$, $t$, $b$, $\eps$, $r$ and $n_0$.

Combining Proposition \ref{secloccorr} with Proposition \ref{compprop},
we see that
\be
	\limsup_{t\to\infty}{\log L(t)\over \log t}
	\leq 
\label{harom}
\ee
$$
	\limsup_{t\to\infty} (\log t)^{-1}
	\log\Bigg[ \cE \exp (-tK\wt\lambda_b^{w(r)})
	 +  \cE \exp\Big( - te^{-\varrho(B+2\beta)G_\Omega}\Big)
	 + \exp(-4Kt)\Bigg]
$$
with $\wt\lambda_b= \wt\lambda_b(B+2\beta)$.
Since the $n_0\to\infty$ limit will always be taken before $\beta\to0$,
the condition $n_0\ge (B/\beta)^{1/2}$ of Proposition \ref{secloccorr}
is satisfied.
The last term in (\ref{harom}) is negligible for small enough $\eps, r$ and
large enough $b$, using (\ref{K1}). 
The estimates on the other two terms are given in the
following propositions:

\begin{proposition}\label{lambdab} For any magnetic field $B>0$
\be
 	\limsup_{n_0\to\infty} \limsup_{r\to0}
	\limsup_{b\to\infty\atop\eps\to0} \limsup_{t\to\infty}
	 (\log t)^{-1}\log \cE \exp \Big(-tK
	\big[\wt\lambda_b(B)\big]^{w(r)}\Big)
	= -\pi \nu ( L^2 + 2B^{-1}) \; .
\label{lambdabeq}
\ee
\end{proposition}

\begin{proposition}\label{ebg} For any magnetic field
$B>0$ 
\be
	\limsup_{n_0\to\infty}\limsup_{r\to0}
	\limsup_{b\to\infty\atop\eps\to0} \limsup_{t\to\infty}
	 (\log t)^{-1}\log 
	\cE \exp\Big( - te^{-BG_\Omega}\Big)
	\leq -{2\pi \nu \over  B } \; .
\label{ebgeq}
\ee
\end{proposition}

Theorem \ref{main} follows from these propositions via (\ref{lapl})
 just by choosing
$\varrho < 2/(2+BL^2)$, using Proposition \ref{lambdab}
with a magnetic field $B+2\beta$ and  Proposition \ref{ebg}
with a magnetic field $\varrho(B+2\beta)$,  and finally letting
$\beta\to0$. $\;\;\;\Box$

{\it Proof of Proposition \ref{lambdab}.} 
Let $\Omega$ be an arbitrary domain and let $B>0$ fixed. 
Let 
$$
	\wh \lambda^{(B)}(\Omega) : = \inf \Bigg\{
	\inf \mbox{Spec} \Big( \frac{1}{2}\big[ (-i\nabla - \wh A)^2
	- B\big] + V_\Omega\Big)_\Omega \; : \; \wh A\in \cA(\Omega)\cap
	C^\infty(\overline{\Omega}), \; \mbox{curl} \,\, \wh A = B
	\; \mbox{on} \; \Omega \Bigg\}
$$
be the smallest eigenvalue of the magnetic Hamiltonian with boundary
potential and Dirichlet boundary conditions on $\Omega$.
We also took the infimum over all possible gauges, which
is unnecessary for simply connected $\Omega$. Here $\cA(\Omega)$ is
the set of real analytic vectorfields on $\Omega$.

In Section \ref{iso} we show the following estimate for
$\wh \lambda^{(B)}(\Omega)$:

\begin{proposition}\label{isoprop} 
For  any $\kappa >0$, $L>0$, $B>0$ and any domain $\Omega$ with volume
 $|\Omega|\ge C(\kappa, L, B)$,
we have 
\be
	 \wh  \lambda^{(B)}
	( \Omega) \ge \exp{ \Big[ - {|\Omega| \over \pi(L^2 + 2B^{-1})}
	(1+\kappa) \Big]} \; .
\label{isoest}
\ee
\end{proposition}

Using this estimate and that $ \wh  \lambda^{(B)}$ is a monotone
function of the domain, the proof of Proposition \ref{lambdab}
is identical to the argument in Section 8 of \cite{E}. $\;\;\;\Box$

{\it Proof of Proposition \ref{ebg}.} We consider 
$$
	\Omega^* := S\setminus \bigcup_{i\in\cG} \overline{B}(x_i, a)
$$
with some fixed $0<a< R$,  we  let $\wt V^* := v\cdot
\1 \Big( x\in \bigcup_{i\in \cG}  \overline{B}(x_i, a)\Big)$
with $v=e^{-(2R)^2/L^2}$
and we let $\wt \lambda^* = \wt\lambda^*(B)$
 be the infimum over $\ua$ of the lowest eigenvalue
of $\frac{1}{2}[ (-i\nabla - A_\ua)^2 - B_\ua] + \wt V^*$.
 These are exactly the set $\Omega$, the potential $\wt V$ and the
eigenvalue $\wt\lambda$ in \cite{E}, but here we use the star  superscript
to distinguish them from their counterparts used in the present paper.

We claim that for any fixed $B>0$, $n_0, R, L, r, \eps, b$
\be
	 \limsup_{t\to\infty}
	 (\log t)^{-1}\log 
	\cE \exp\Big( - te^{-BG_\Omega}\Big)
	\leq 
	 \limsup_{t\to\infty}
	 (\log t)^{-1}\log 
	\cE \exp\Big( - te^{-BG_{\Omega^*}}\Big) \; .
\label{gomcomp}
\ee
For the proof, we define 
$$
	\Omega^\# := S\setminus \bigcup_{i\in\cG} \overline{B}(x_i, 2R) 
	\; .
$$
Clearly
 $g_{\Omega}(x) = g_{\Omega^\#}(x)$ for any $x \not\in \bigcup_{i\in \cG}
\overline{B}(x_i, R)$, while $g_{\Omega}(x) \leq c_9(R)$
if $x\in \overline{B}(x_i, R)$ for some $i\in \cG$. Hence
$G_\Omega \leq G_{\Omega^\#} + c_9(R) \leq G_{\Omega^*} + c_9(R)$,
where the second inequality follows from $\Omega^\# \subset \Omega^*$.

\bigskip

 We then recall that Section 8 of \cite{E}, from (8.1) through (8.8)
actually gave the following bound (there $B$ was replaced by $B+2\beta$):
\be
	\limsup_{n_0\to\infty} \limsup_{r\to0}
	\limsup_{b\to\infty\atop\eps\to0} \limsup_{t\to\infty}
	 (\log t)^{-1}\log 
	\cE e^{- tN\wt\lambda^*(B)}
	\leq -{2\pi\nu\over B} \; .
\label{sec8}
\ee
for any function $N$ satisfying 
$$
	 \limsup_{r\to0}
	\limsup_{b\to\infty\atop\eps\to0}
	\limsup_{t\to\infty} {-\log N\over \log t} =0 \; .
$$
Using Lemma 7.7 of \cite{E}, stating that $\wt\lambda^*$ is smaller than
$e^{-BG_{\Omega^*}}$ modulo negligible factors,
 we easily obtain (\ref{ebgeq}) from 
(\ref{gomcomp}) and (\ref{sec8}). $\;\;\;\Box$

\section{Estimate on the magnetic eigenvalue
with a boundary potential}\label{iso}

In this section we prove Proposition \ref{isoprop}. 
Let $D$ be the disk of radius $R_\Omega := \pi^{-1/2} |\Omega|^{1/2}$
centered at the origin. For any function $a(r)$ with
\be
	0 \leq 2\pi a(r) r \leq B\pi r^2 \; , \qquad \mbox{for all}\quad
	0 \leq r \leq R_\Omega \; ,
\label{gaugecomp}
\ee
we define a radial gauge $A_{rad}(x)= a(r)\pmatrix{-\sin\theta\cr \cos\theta}$
(in polar coordinates, $x= re^{i\theta}$) that generates the
radial magnetic field $B_{rad}(x) = \mbox{curl} \,\, A_{rad}(x)
= a'(r) + r^{-1} a(r)$.
Condition (\ref{gaugecomp}) requires the flux of the magnetic field
$B_{rad}$ to be not smaller than that of the constant $B$ field
on all concentric disks $B(0,r)$.

Let $\cH$ and $\cH_D$ be the Hilbert spaces of radially symmetric
$H^1(\bR^2)$ and $H^1_0(D)$ functions, respectively. Let
$$
	H(a) := \frac{1}{2}\Big[ (-i\nabla - A_{rad})^2 - B_{rad}\Big]
$$
be defined on $\cH_D$, and let $\lambda(a)$ be its lowest eigenvalue.
It is easy to see that the corresponding eigenfunction can be chosen
nonnegative. In fact
\be
	\langle f, H(a) f\rangle \ge \langle |f|, H(a) |f|\rangle
	=  \Big\langle |f|, {1\over 2}
	(-\Delta + a^2 - B_{rad}) |f|\Big\rangle 
	\qquad f\in \cH_D \; .
\label{radcomp}
\ee
Proposition 2.1 of \cite{E-1996} states that the lowest magnetic
Dirichlet eigenvalue on $\Omega$ with a constant magnetic field $B$
is minorized by $\lambda(a)$ for some $a(r)$ that satisfies
 (\ref{gaugecomp}).

For any nonnegative function $\psi$
we denote its symmetric rearrangement by $\psi^*$, i.e., $\psi^*$
is the unique radial function with the property that $|\{ \psi \ge c \}|
= |\{ \psi^* \ge c \}|$ for any $c$.
It is not stated explicitly in
\cite{E-1996}, but actually
 the proof of  Proposition 2.1 in \cite{E-1996} gives 
the following  result from which the comparison
of the eigenvalues has been derived. 

\begin{proposition}\label{freeest}
 Let $\wh A \in \cA(\Omega)\cap C^\infty(\overline\Omega)$,
$\mbox{curl} \,\, \wh A = B$ on $\Omega$ and let
$\wh H ={1\over 2}\Big[ (-i\nabla - \wh A)^2 - B \Big]$. Then for any function
$f\in H_0^1(\Omega)$ there exists a  function $a(r)$
that satisfies (\ref{gaugecomp}) such that
$$
	\langle f, \wh H f \rangle
	\ge \langle |f|^*,  H(a) |f|^* \rangle  \; .
$$
\end{proposition}
For the proof one only has to notice that the radial trial
function $q(r)$
defined on p. 289 of \cite{E-1996} as $q(r)= \Lambda^{-1}(h^*(r))$ 
is actually the symmetric rearrangement of $\psi = |f|$ since
 $h=\Lambda(\psi)$ and $\Lambda$ is strictly monotone. $\;\;\Box$

\bigskip

Now we include the boundary potential $V_\Omega$ (see (\ref{bp})).
 We replace $V_\Omega$
by 
$$
	\wh V_\Omega := {1\over \pi L^2} \cdot 
	\1_{\Omega^c} \star e^{-| \, \cdot\,  |^2/L^2}
$$
where $ \1_{\Omega^c}$ is the characteristic function of $\Omega^c$
and $\star$ denotes the convolution.
It is straightforward that $\wh V_\Omega \leq V_\Omega$
on $\Omega$ and $\wh V_\Omega\leq 1$ everywhere.

By the Riesz rearrangement inequality,
$$
	\langle f, (1-\wh V_\Omega) f \rangle 
	= {1\over \pi L^2}
	 \int\!\!\!\int  |f(x)|^2 e^{-|x-y|^2/L^2} \1_\Omega(y) \rd x \rd y
	\leq {1\over \pi L^2}
	\int\!\!\!
	\int \big( |f|^2 \big)^*(x) e^{-|x-y|^2/L^2} \1_{D}(y) 
	\rd x \rd y \; ,
$$
where the disk $D$ is the symmetric rearrangement of the set $\Omega$.
A simple estimate yields
$$
	\langle f, \wh V_\Omega f \rangle \ge	
	\langle |f|^* , W_\eta |f|^* \rangle
$$
for any $f\in L^2(\Omega)$ and any $\eta>0$ with
$$
	W_\eta(x) 
	:= {1\over 2}\exp{\Big( - (1+\eta^{-1}) - 
	{(R_\Omega-|x|)^2\over L^2}(1+\eta)
	\Big)}\; , \qquad |x| \leq R_\Omega \; .
$$
{F}rom these estimates and Proposition \ref{freeest} we conclude
that there exists a radial function $a(r)$, satisfying 
(\ref{gaugecomp}), such that
$$
	\wh\lambda^{(B)}(\Omega) \ge \inf \, \mbox{Spec}
	\Big( H(a) + W_\eta \Big) \qquad \mbox{on} \quad \cH_D \; .
$$

Next we claim that if $a_1(r) \leq a_2(r)$ satisfy (\ref{gaugecomp}),
then 
$$
	\inf \, \mbox{Spec}
	\Big( H(a_1) + W_\eta \Big)
	\ge \inf \, \mbox{Spec}
	\Big( H(a_2) + W_\eta \Big) \qquad \mbox{on} \;\; \cH_D\; .
$$
This is proven exactly as Lemma 3.1 in \cite{E-1996}. It is easy
to check that the inclusion of a bounded
nonnegative radial
potential $W_\eta$ does not alter the trial function argument.

Therefore $H(a)+W_\eta$ has the lowest eigenvalue if $a(r)= Br/2$,
i.e., in case of the constant field.
Using (\ref{radcomp}), this eigenvalue is the same as the lowest
eigenvalue, $\lambda_{\eta}$, of
$$
	H_\eta := H_{osc} + W_\eta \; ,\qquad \mbox{with}\quad
	 H_{osc} := {1\over 2} \Big[ -\Delta + {Bx^2\over 4} - B\Big] 
	\qquad \mbox{on} \;\; \cH_D\; .
$$

Let $\varphi_0 (x)= \sqrt{B\over 2\pi} e^{-Bx^2/4}$ span the
kernel of the harmonic oscillator
$H_{osc}$ on $\cH$,
and let $P: = |\varphi_0\rangle \langle \varphi_0|$ be the
projection onto this kernel. It is well known that $H_{osc}$
has a gap of size $B$ above zero on $\cH$. We can estimate $\lambda_{\eta}$
by decomposing the eigenfunction $f\in \cH_D$ as $f= Pf + (I-P)f$:
\be
	\lambda_{\eta} = \Big\langle (I-P)f, H_{osc} (I-P)f\Big\rangle
	+ \langle f, W_\eta f\rangle
	\ge B \| (I-P)f\|^2 + \langle f, W_\eta f\rangle \; .
\label{lamalul}
\ee
Furthermore,
$$
	\lambda_{\eta} \ge \int_D  W_\eta |f|^2
	\ge {1\over 2} \int_D  W_\eta |P f|^2- 2 
	\int_D  W_\eta |(I-P) f|^2	
	\ge {1\over 2} \int_D  W_\eta |P f|^2-  \lambda_{\eta} B^{-1} \; ,
$$
using (\ref{lamalul}) and $W_\eta\leq 1/2$. Hence
\be
	\lambda_{\eta} \ge {B\over 2(B+1)}  \int_D  W_\eta |Pf|^2 \; .
\label{cle}
\ee
Since $ \| Pf\|^2 + \|(I-P)f\|^2= \| f \|^2 =1$ and $\| (I-P)f\|^2
\leq \lambda_{\eta} B^{-1}$ from (\ref{lamalul}), we have $\|Pf \|^2 =
|\langle f, \varphi_0\rangle|^2
 \ge 1-\lambda_{\eta} B^{-1}$. We can assume that $\lambda_{\eta}
< B/2$,  otherwise 
Proposition \ref{isoprop} is trivial.
Hence
\be
	\lambda_{\eta} \ge {B\over 4(B+1)} 
	 \int_D  W_\eta |\varphi_0|^2
	= {B^2\over 16\pi ( B+1)} e^{-(1+\eta^{-1})}\int_D
	e^{-(1+\eta)(R_\Omega -|x|)^2/L^2} e^{- Bx^2/2} \rd x
\label{int}
\ee
$$
	\ge C(B, L)
	\exp{\Big[ - (1+\eta^{-1})- {R_\Omega^2\over L^2 + 2B^{-1}}(1+\eta)
	\Big]} \; .	
$$
 {F}rom this bound,
Proposition \ref{isoprop} easily follows. $\,\,\,\Box$

{\it Remark:} From the integration (\ref{int}) one can see
 the interplay between
the Gaussian eigenfunction $\varphi_0$ and the Gaussian 
potential $W_\eta$. In particular, the main contribution comes
from the intermediate regime around $|x| \approx {2\over BL^2+2}\,
R_\Omega$.

\end{document}